\documentstyle[11pt,icfa,epsfig]{article}

\title{Status Report on the double-$\beta$ decay experiment NEMO-3}

\author{NEMO Collaboration\\[0.6cm]
{\footnotesize{\em CENBG IN2P3-CNRS et Universit\'e de Bordeaux, France}} \\
{\footnotesize{\em CFR, CNRS, Gif-sur-Yvette, France}} \\
{\footnotesize{\em FNSPE, Prague University, Czech Republic}} \\ 
{\footnotesize{\em INEEL, Idaho Falls, USA}} \\ 
{\footnotesize{\em IReS, IN2P3-CNRS et Universit\'e de Strasbourg, France}} \\ 
{\footnotesize{\em ITEP, Moscow, Russia}} \\ 
{\footnotesize{\em JINR, Dubna, Russia}} \\ 
{\footnotesize{\em Jyvaskyla University, Finland}} \\ 
{\footnotesize{\em LAL, IN2P3-CNRS et Universit\'e Paris-Sud, France}} \\ 
{\footnotesize{\em LPC, IN2P3-CNRS et Universit\'e de Caen, France}} \\ 
{\footnotesize{\em Mount Holyoke College, USA}} \\[0.6cm]
Contributed Paper for the XIX International Conference on Neutrino Physics and Astrophysics, Neutrino 2000, Sudbury, Canada, June 16-21, 2000, presented by X. Sarazin \\
and for the XXX International Conference on High Energy Physics, \\
ICHEP2000, Osaka, Japan, July 27 - August 2, 2000, presented by D. Lalanne.}

\begin{document}

\maketitle

\begin{abstract}

The NEMO collaboration is presently mounting the NEMO-3 detector in the Fr\'ejus Underground Laboratory. This detector, which will be completed by the end of the year 2000, is devoted to the search of neutrinoless double beta decay with various isotopes. Much attention has been focused on $^{100}$Mo and  $^{82}$Se with their large $Q_{\beta\beta}$-values. The detector is based on the direct detection of the two electrons by a tracking device and on the measurement of their energies by a calorimeter. The aim of the experiment is to have a sensitivity for the effective neutrino mass on the order of 0.1~eV. The status and the expected performance of the NEMO-3 detector for both internal and external background rejections and for signal detection are presented. 

\end{abstract}

\section{Introduction}

Several strong indications in favor of neutrino masses and mixing have been observed in atmospheric and solar neutrinos. However, direct detection of neutrino masses has not been measured. The most stringent upper limit obtained by tritium beta decay is $m_{\nu}<2.8$~eV (95\% C.L.) \cite{mainz}. 
Another fundamental question of neutrino physics is the nature of massive neutrinos. Are massive neutrinos Dirac particles or neutral Majorana particles having all lepton numbers equal to zero? 
The neutrinoless double beta decay $\beta\beta(0\nu)$ which is a process beyond the electroweak Standard Model, is the only way to prove the existence of Majorana neutrinos. In some phenomenologically viable neutrino scenarios, the effective Majorana neutrino mass $\langle m_{\nu} \rangle$ can be 0.1 eV (in a three-neutrino scenario with two mass-degenerate neutrinos) or even as large as 1 eV (in a four-neutrino scenario which accomodates all the oscillation measurements) \cite{bilenki}. 

To date, the most stringent limit on the $\beta\beta(0\nu)$ half-life is obtained in the $^{76}$Ge Heidelberg-Moscow experiment \cite{laura}: 
\begin{equation} 
T_{1/2}^{0\nu} > 1.6 \; 10^{25} \mbox{ yr   (90\% C.L.)}
\end{equation} 
 From this limit, an upper limit on $\langle m_{\nu} \rangle$ can be inferred with the relation: 
\begin{equation} 
(T_{1/2}^{0\nu})^{-1} = \left(\frac{\langle m_{\nu} \rangle}{m_e}\right)^2 \times |M_{0\nu}|^2 \times F_{0\nu}
\end{equation} 
where $M_{0\nu}$ is the nuclear matrix element of the relevant isotope 
and $F_{0\nu}$ is the phase-space factor. \\
Calculations of $M_{0\nu}$ have unfortunately large theoretical uncertainties. 
Depending on the calculation of $M_{0\nu}$, one obtains limits on $\langle m_{\nu} \rangle$ ranging from 0.4 eV to 1 eV \cite{laura}. The limit  $\langle m_{\nu} \rangle <$ 1 eV is obtained by using calculations performed in the framework of the Shell Model \cite{caurier}. 
$F_{0\nu}$ is analytically calculable and is proportional to $Q_{\beta\beta}^5$ ($Q_{\beta\beta}=2040 keV$ for $^{76}$Ge) . Therefor to improve the sensitivity of a double-$\beta$ decay experiment, an isotope with a larger $Q_{\beta\beta}$ seems to be preferable in order to get a larger $F_{0\nu}$, but also to reduce the background in the search for a $\beta\beta0\nu$ signal.

The aim of the NEMO-3 detector, which will operate in the Fr\'ejus Underground Laboratory, referred to as the Laboratoire Souterrain de Modane (LSM), is to search for $\beta\beta(0\nu)$ with various isotopes with large Q$_{\beta\beta}$ values. The detector is able to accomodate at least 10 kg of double beta decay isotopes.
Much attention has been focused on $^{100}$Mo ($Q_{\beta\beta}$ = 3034 keV), $^{82}$Se ($Q_{\beta\beta}$ = 2995 keV) and $^{116}$Cd ($Q_{\beta\beta}$ = 2802 keV).

\section{The NEMO-3 detector}

\begin{figure}[htb]
\begin{center}
\mbox{
\epsfxsize=14cm
\epsfbox{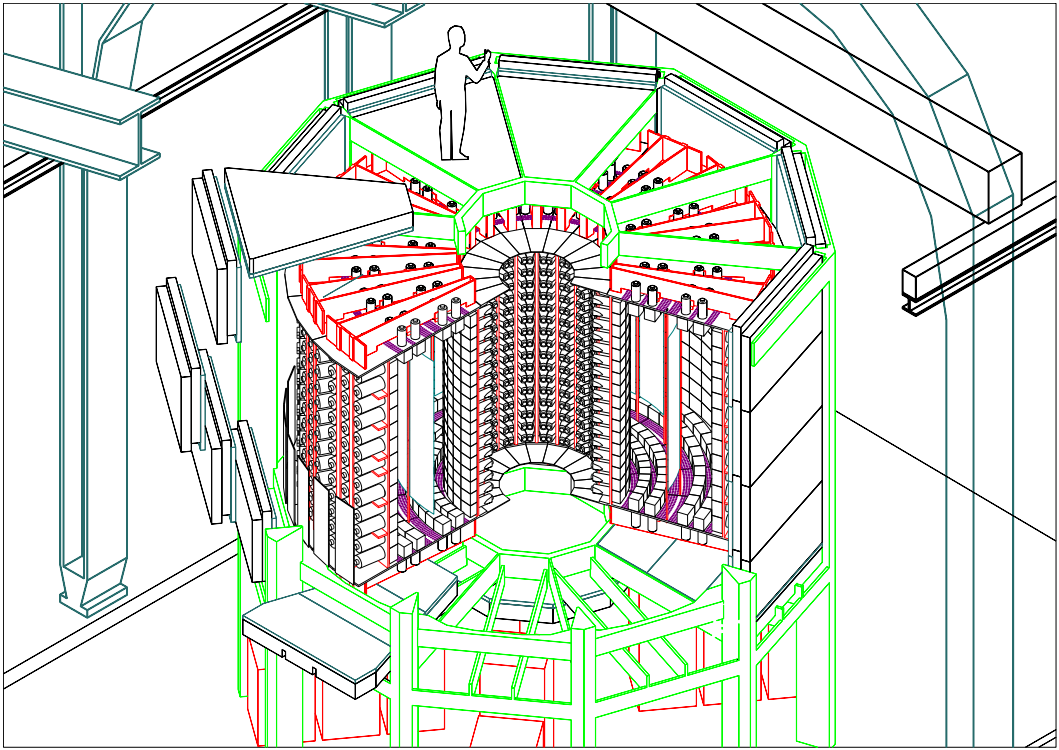}}
\caption{Layout of the NEMO-3 detector.}
\label{fig:nemo3}
\end{center}
\end{figure}

The NEMO-3 experiment is based on the direct detection of the two electrons by a tracking device and on the measurement of their energies by a calorimeter. 
The NEMO-3 detector, as shown in the Figure~\ref{fig:nemo3}, is similar in function to the earlier prototype NEMO-2 \cite{nemo2}.  \\
The detector is cylindrical in design and divided into 20 equals sectors. Thin ($\sim 50 \mu m$) source foils are fixed vertically between two concentric cylindrical tracking volumes composed of open octagonal drift cells, 270 cm long, operating in Geiger mode. In order to minimize multiple scattering effects, the tracking volume is filled with a mixture of helium gas and 4\% ethyl alcohol. The wire chamber provides three-dimensional tracking. The tracking volume is covered with calorimeters made of large blocks of plastic scintillators coupled to very low radioactivity 3'' and 5'' PMTs. The finished detector contains 6180 drift-Geiger cells and 1940 scintillators. 

A solenoid surrounding the detector produces a magnetic field of 30 Gauss in order to recognize ($e^+e^-$) pair production events in the source foils. An external shield, in the form of 20 cm thick low radioactivity iron, covers the detector to reduce $\gamma$-rays and thermal neutron fluxes. Outside of this shield, an additional shield is added to thermalize fast neutrons. 

\section{Current Status of Construction}

The construction of the 20 sectors of the NEMO-3 detector has been completed. Currently, 12 sectors are  in the Underground Laboratory and 6 of them are equipped with their source foils and mounted on the detector frame. The detector will be completed by the end of the year 2000. 

The energy resolution of each scintillator block has been measured with a 1 MeV electron spectrometer during the construction of the calorimeter. The energy resolution is $\sigma(E)/E=5.6\%$ at 1 MeV which is lower than the energy resolution of 7\% at 1 MeV specified in the detector's proposal.

The double-$\beta$ decay isotopes which are being mounted in the detector are the following: 7 kg of $^{100}$Mo (corresponding to 12 sectors), 1 kg of $^{82}$Se (2.3 sectors), 0.6 kg of $^{116}$Cd (1 sector), 0.7 kg of $^{130}$Te (1.8 sectors), 50 g of $^{150}$Nd, 16 g of $^{96}$Zr and 8 g of $^{48}$Ca. Also, 2.7 sectors are devoted to external background measurements: 1 sector is equipped with an ultra-pure copper foil and 1.7 sectors with 0.9 kg of $^{nat}$TeO$_2$. 
To date, $^{82}$Se, $^{116}$Cd, $^{nat}$TeO$_2$ and the copper foils are mounted. The choice of Cu and $^{nat}$TeO$_2$ is explained below. 
We are now starting to mount the $^{100}$Mo sources.

\begin{figure}[htb]
\begin{center}
\mbox{
\epsfxsize=12cm
\epsfysize=12cm
\epsfbox{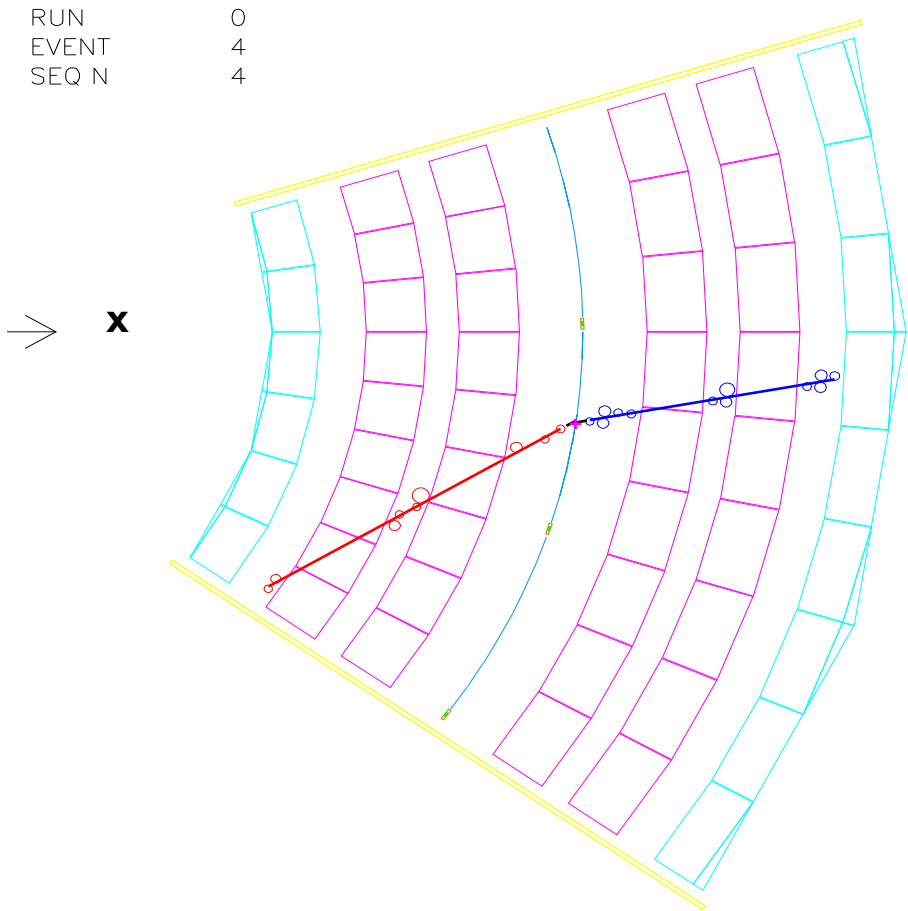}}
\caption{Transverse view of two Geiger tracks measured with the first 3 sectors of NEMO-3 running in the Fr\'ejus Underground Laboratory. Small open circles are the activated Geiger cells, the different radii correspond to the transverse drift distance to the anodic wire.}
\label{fig:event1}
\end{center}
\end{figure}

\begin{figure}[htb]
\begin{center}
\mbox{
\epsfxsize=12cm
\epsfysize=12cm
\epsfbox{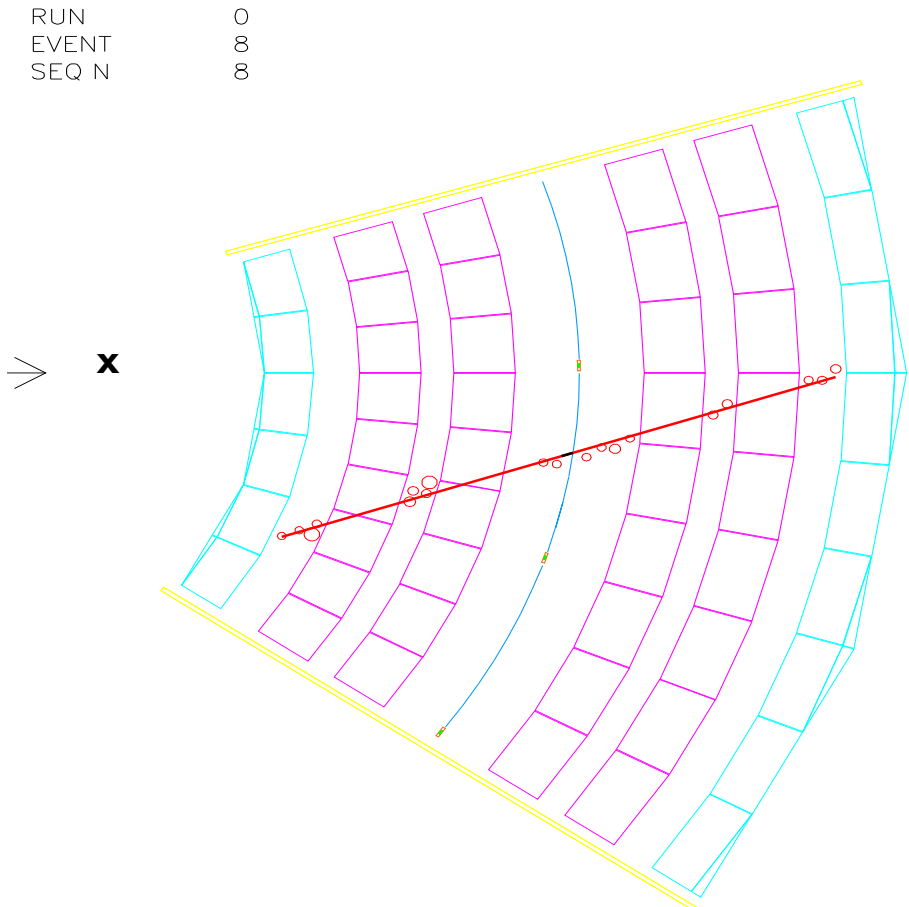}}
\caption{Another transverse view of two Geiger tracks measured with the first 3 sectors of NEMO-3 running in the Fr\'ejus Underground Laboratory.}
\label{fig:event2}
\end{center}
\end{figure}

Three sectors installed on the detector frame have been succesfully running since the end of April 2000 (without a magnetic field and an external shield). The NEMO collaboration has decided to start operating with these 3 sectors in order to test the tracking and calorimeter parts of the detector. The wire chamber and the PMTs coupled to the scintillators are running well and only 0.3\% of Geiger cells are out-of-order. Geiger $\beta$ tracks obtained with these 3 sectors and with the finalized NEMO-3 trigger and acquisition system, are shown in Figure~\ref{fig:event1} and \ref{fig:event2}.

\section{Expected background}

There are three origins of expected background which can occur in this search for a $\beta\beta0\nu$ signal around 3 MeV. 
The first  background comes from the beta decays of $^{214}$Bi (Q$_{\beta} =$ 3.2 MeV) and $^{208}$Tl (Q$_{\beta} =$ 5.0 MeV) which are present in the source, from the Uranium and Thorium decay chains. They can mimic $\beta\beta$ events by $\beta$ emission followed by M$\ddot{o}$ller effect or by a $\beta-\gamma$ cascade followed by a Compton interaction. Thus, the experiment requires ultra-pure enriched $\beta\beta$ isotopes. 
A second origin of $\beta\beta0\nu$ background is due to high energy gamma rays ($>$ 2.6 MeV) interacting with the source foil. Their origin is from  neutron captures occuring inside the detector. The interactions of these gammas in the foil can lead to 2 electrons by $e^+e^-$ pair creation, double Compton scattering or Compton followed by M$\ddot{o}$ller scattering. 
Finally, given the energy resolution, the ultimate background is the tail of the $\beta\beta2\nu$ decay distribution. It defines the half-life limits to which the $\beta\beta0\nu$ can be studied.

\subsection{Radiopurity of the sources in  $^{214}$Bi and $^{208}$Tl}

\subsubsection{$^{100}$Mo source}

Maximum levels of  $^{214}$Bi and $^{208}$Tl contamination in the source have been calculated to insure that $\beta\beta2\nu$ is the limiting background. These limits  are $^{214}$Bi $<$ 0.3 mBq/kg and $^{208}$Tl $<$ 0.02 mBq/kg. 
These activities in $^{214}$Bi and $^{208}$Tl correspond to a level of $2 \; 10^{-11}$~g/g in $^{238}$U and $10^{-11}$~g/g in $^{232}$Th respectively when we assume the natural radioactive families of $^{238}$U and $^{232}$Th are in equilibrium.
To reach these specifications, two methods have been developed to purify the enriched Molybdenum isotope.

The first method developed by ITEP (Moscow, Russia), is a purification by local melting of solid Mo with an electron beam and drawing a monocrystal from the liquid portion. One gets an ultra-pure $^{100}$Mo monocrystal. The crystal is then rolled into a metallic foil for use in the detector. Much attention has been focused on this rolling process. To date 0.5 kg of foil has been produced and no contaminant activity have been measured with HP-Ge in the LSM.

The second purification method is a chemical process done at INEEL (Idaho, USA) which leaves the Mo in a powder form that is then used to produce foils with a binding paste and mylar strips which have been etched with an ion beam and a chemical process. To date 3 kg of  $^{100}$Mo have been purified and 2~kg more are being processed and will be ready towards the end of September 2000. No activity has been observed in the purified $^{100}$Mo after 1 month of HP-Ge measurements in the LSM and the most stringent limits obtained for radiopurities are $^{214}$Bi $<$ 0.2 mBq/kg and $^{208}$Tl $<$ 0.05 mBq/kg. The radiopurity in $^{214}$Bi is already better than the design specifications. The task of measuring the required limits for $^{208}$Tl is beyond the practical measuring limits of the HP-Ge detectors in the LSM. However, the chemical extraction factors defined as the ratio of  contamination before and after purification were measured with a $^{nat}$Mo sample. Applying the $^{208}$Tl extraction factor to the $^{208}$Tl activity measured in the $^{100}$Mo before purification, one obtains after purification an expected level in $^{208}$Tl of 0.01 mBq/kg which is again lower than the design specifications. 

\subsubsection{$^{82}$Se source}

Some low activities in $^{214}$Bi and $^{208}$Tl have been measured in the 1 kg $^{82}$Se source foils with HP-Ge studies. The activities are 1.2 $\pm$ 0.5 mBq/kg in $^{214}$Bi and 0.4 $\pm$ 0.1 mBq/kg in $^{208}$Tl. This corresponds to an expected background of 0.2 events/yr/kg from $^{214}$Bi and 1 event/yr/kg from $^{208}$Tl.\\
The same contamination had been measured with $^{82}$Se foils used in the NEMO-2 prototype and contaminants were found to be concentrated in small ``hot-spots'' and rejected in the analysis thanks to the tracking device \cite{selenium_nemo2}. We believe that the contamination in these $^{82}$Se foils is identical and will be suppressed by software analysis.

\subsection{External background from neutrons and $\gamma$-rays}

The effect of neutrons and $\gamma$-rays on the background in the $\beta\beta0\nu$ energy region was studied for 10,700 hours of live time with the NEMO-2 prototype \cite{neutrons}. Various shields and measuremements with a neutron source were used to identify the different components.

This study has shown that there is no contribution from thermal neutrons which are stopped in a few centimeters of the iron shielding but that the dominating background is due to fast neutrons ($>$ 1 MeV) from the laboratory. 
Fast neutrons going through the iron shielding, are thermalized in the plastic scintillators and then captured in copper, iron or hydrogen inside the detector.
To compare the data and Monte Carlo calculations, a study required the development of an interface between GEANT/MICAP and a new library for $\gamma$-ray emission after capture or inelastic scattering of neutrons. Good agreement was obtained between the experiments and simulations.

It was demonstrated with the neutron simulations for NEMO-3 that an appropriate neutron shield (like paraffin) and a 30 Gauss magnetic shield will make the neutron background negligible \cite{neutrons}.

\subsection{Radiopurity of the detector}

Additionally, the components of the detector have to be ultra-pure in $^{214}$Bi, $^{208}$Tl and $^{40}$K to have a low background in the $\beta\beta2\nu$ energy spectrum.
This is required to not only measure the $\beta\beta2\nu$ period with high accuracy but also to see any distortions in the $\beta\beta2\nu$ spectrum due to Majoron emission. 
Finally, the high radiopurity is required so that we can measure the $e\gamma$ and $e\gamma\gamma$ events which identify the Tl activity in the source.

The activities of all materials used in the detector were measured with HP-Ge detectors in the LSM  or at the CENBG laboratory in Bordeaux (France). This exhausting examination of samples, corresponding to about 1000  measurements, reasulted in the rejection of numerous glues, plastics, and metals. Activities in $^{214}$Bi, $^{208}$Tl and  $^{40}$K, of the main components of the detector are listed in Table~\ref{tab:activities}.

As expected, the radioactive contamination in the detector is dominated by the low radioactivity glass in the PMTs. The activity of these PMTs are three orders of magnitude below standard PMT levels. 
With a total activity of 300~Bq for $^{214}$Bi and 18~Bq for $^{208}$Tl in the 600~kg of PMTs, the expected signal-to-background ratio ($S/B$) in the integrated $\beta\beta2\nu$ energy spectrum is $S/B \sim 400$ from $^{214}$Bi  and $S/B \sim 900$ from $^{208}$Tl with 7~kg of $^{100}$Mo (T$_{1/2}(\beta\beta2\nu) = 0.95 \; 10^{19}$y). This ratio becomes about 10 times smaller with $^{82}$Se since its $\beta\beta2\nu$ half-life is about 10 times larger (T$_{1/2}(\beta\beta2\nu) = 0.8 \; 10^{20}$y). 

Activities of all other components are under our measurement sensitivity and negligeable compare to the PMTs.

\begin{table}[h]
\begin{center}
\begin{tabular}{c|c|cccc}
\hline
           &             &  Total   & Activity   & (in Bq)    &           \\
Components & Weight (kg) & $^{40}$K & $^{214}$Bi & $^{208}$Tl & $^{60}$Co \\
\hline
 PMTs  & 600 & 830 & 300 & 18 & \\        
 scintil. & 5,000 & $<$100 & $<$0.7 & $<$0.3 & 1.8 $\pm$ 0.4\\
 copper & 25,000 & $<$125 & $<$25 & $<$10 & $<$6 \\
 petals iron & 10,000 & $<$50 & $<$6 & $<$8 & 17 $\pm$ 4 \\
 $\mu$ metal & 2,000 & $<$17 & $<$2 & 2.0 $\pm$ 0.7 & 4.3 $\pm$ 0.7 \\
 wires & 1.7 & $<$8.10$^{-3}$ & $<$10$^{-3}$ & $<$6.10$^{-4}$ & 10$^{-2}$ \\
 shield. iron & 180,000 & $<$3000 & $<$300 & $<$300 & 300 $\pm$ 100 \\
\hline 
\end{tabular}
\caption{ Total activities (in Bq) for the main components of the NEMO-3 detector, measured with HP-Ge detectors in the Fr\'ejus Underground Laboratory.}
\label{tab:activities}
\end{center}
\end{table}

\subsection{$^{nat}$TeO$_2$ and Copper foils to measure external background}

Foils of $^{nat}$TeO$_2$  inserted into the NEMO-3 detector allow one to measure the external background for $^{100}$Mo. The effective $Z$ of these foils is nearly the same as that of molybdenum foils. This is useful because the external $\gamma$-ray background can give rise to pair production, double Compton scattering, or Compton-M$\ddot{o}$ller which are all proportional to $Z^2$. Thus, the background for $^{100}$Mo and $^{nat}$TeO$_2$ foils should give rise to similar event rates. However, $^{nat}$TeO$_2$, which is 34.5\% $^{130}$TeO$_2$, produces no $\beta\beta$ pairs in the energy region above the $Q_{\beta\beta}$-value of $^{130}$TeO$_2$ (2.53 MeV), so a background subtraction is possible for $^{100}$Mo foils given the spectrum of $^{nat}$TeO$_2$. The copper foils provide a similar study for a smaller value of $Z$.

\subsection{Number of background events in the $\beta\beta0\nu$ energy region}

The expected numbers of background events, in the energy range 2.8 to 3.2 MeV around the $\beta\beta0\nu$ signal peak are summarized in Table~\ref{tab:expect_bkg} for $^{100}$Mo and $^{82}$Se. 

\begin{table}[h]
\begin{center}
\begin{tabular}{c|c|c}
\hline
            & $^{100}$Mo  &  $^{82}$Se    \\
            & events/yr/kg &  events/yr/kg  \\
\hline
 $^{214}$Bi &   $<$ 0.03  &   negl.           \\        
 $^{208}$Tl &   $<$ 0.04  &   negl.           \\        
 $\beta\beta2\nu$  &   0.11     &  0.01       \\        
 External neutrons &  $<$ 0.01  &  $<$ 0.01   \\        
\hline
 TOTAL &   $<$ 0.18     & 0.01 \\        
\hline 
\end{tabular}
\caption{ Expected number of background events, in the energy window 2.8 to 3.2 MeV, per year per kg. For $^{82}$Se, it is believed that the background from $^{214}$Bi and $^{208}$Tl will be limited to ``hot-spots'' (see text).}
\label{tab:expect_bkg}
\end{center}
\end{table}

\section{Expected sensitivity of NEMO-3}

The sensitivity that the NEMO-3 detector will reached after 5 years of data collection, has been calculated with 7~kg of $^{100}$Mo and 1~kg of $^{82}$Se. 
After 5 years, in the energy window 2.8 to 3.2 MeV, a total of 6 background events are expected with 7~kg of $^{100}$Mo and no background events are expected with 1~kg of $^{82}$Se. 
The $\beta\beta0\nu$ detection efficiency in the same energy window, 2.8 to 3.2~MeV, is $\epsilon(\beta\beta0\nu) = 14\%$. 
The expected sensitivities are summarized in Table~\ref{tab:sensitivity}. 

\newpage

\begin{table}[h]
\begin{center}
\begin{tabular}{c|c|c}
\hline
    & 7 kg $^{100}$Mo  &  1 kg $^{82}$Se  \\
\hline
Number of events & 6 background events expected & 0 background events expected \\
in the energy window   &  6 events observed           &  0 events observed \\
2.8 to 3.2 MeV        &  5 $\beta\beta0\nu$ excluded &  2.5 $\beta\beta0\nu$ excluded\\
\hline 
$T_{1/2}^{0\nu}$    &  $>$ 4. 10$^{24}$ yr         & $>$ 1.5 10$^{24}$ yr \\
$\langle m_{\nu} \rangle$ & $<$ 0.25 - 0.7 eV      & $<$ 0.6 - 1.2 eV     \\
\hline 
\end{tabular}
\caption{ Expected sensitivity (90\% C.L.) for NEMO-3 after 5 years of data with 7 kg of $^{100}$Mo and 1 kg of $^{82}$Se (the number of events are given in the energy window 2.8 to 3.2 MeV around the $\beta\beta0\nu$ signal peak).}
\label{tab:sensitivity}
\end{center}
\end{table}

\newpage

\end{document}